\documentclass[runningheads]{llncs}
\usepackage[T1]{fontenc}
\usepackage{graphicx,verbatim}
\usepackage{graphicx} 
\usepackage{color,soul}
\usepackage{amsmath}
\usepackage{caption}
\begin{document}
\title{Explainable AI for Automated User-specific Feedback in Surgical Skill Acquisition}
%
\titlerunning{XAI for Automated Feedback in Surgical Skill Acquisition}

\author{
Catalina Gomez \inst{1}\thanks{Equal contribution}\and
Lalithkumar Seenivasan\inst{1\star}\and
Xinrui Zou\inst{1\star}\and
Jeewoo Yoon\inst{1} \and
Sirui Chu\inst{1} \and
Ariel Leong\inst{2} \and
Patrick Kramer\inst{2} \and
Yu-Chun Ku\inst{1} \and
Jose L. Porras\inst{2} \and
Alejandro Martin-Gomez\inst{3} \and
Masaru Ishii\inst{2} \and
Mathias Unberath\inst{1} 
}

\authorrunning{Gomez et al.}
%
\institute{Johns Hopkins University, Baltimore, MD, USA \and
Johns Hopkins Medical Institutions, Baltimore, MD, USA\and
University of Arkansas, Fayetteville, AR, USA\\ 
\email{\{cgomezc1,unberath\}@jhu.edu}
}

\maketitle              
\begin{abstract}

Traditional surgical skill acquisition relies heavily on expert feedback, yet direct access is limited by faculty availability and variability in subjective assessments. While trainees can practice independently, the lack of personalized, objective, and quantitative feedback reduces the effectiveness of self-directed learning. Recent advances in computer vision and machine learning have enabled automated surgical skill assessment, demonstrating the feasibility of automatic competency evaluation. However, it is unclear whether such Artificial Intelligence (AI)-driven feedback can contribute to skill acquisition. Here, we examine the effectiveness of explainable AI (XAI)–generated feedback in surgical training through a human-AI study. We create a simulation-based training framework that utilizes XAI to analyze videos and extract surgical skill proxies related to primitive actions. Our intervention provides automated, user-specific feedback by comparing trainee performance to expert benchmarks and highlighting deviations from optimal execution through understandable proxies for actionable guidance. In a prospective user study with medical students, we compare the impact of XAI-guided feedback against traditional video-based coaching on task outcomes, cognitive load, and trainees’ perceptions of AI-assisted learning. Results showed improved cognitive load and confidence post-intervention. While no differences emerged between the two feedback types in reducing performance gaps or practice adjustments, trends in the XAI group revealed desirable effects where participants more closely mimicked expert practice. This work encourages the study of explainable AI in surgical education and the development of data-driven, adaptive feedback mechanisms that could transform learning experiences and competency assessment.


\end{abstract}

\section{Introduction}
Surgical education faces critical constraints as limited faculty availability and persistent variability in subjective skill assessments undermine consistent, high-quality training~\cite{babineau2004cost}. The Accreditation Council for Graduate Medical Education's 80-hour workweek restriction, while improving the work-life balance, has reduced the time available for hands-on skill development~\cite{wilson2003new}. With fewer opportunities for deliberate practice, surgical education must become more efficient without compromising competency. 
These challenges highlight the urgent need for scalable, objective feedback mechanisms that provide trainees with timely and actionable insights~\cite{scott2006patient}.

Alternative to traditional expert-based evaluation, video-based assessment has emerged as a promising tool for objectively evaluating surgical skills~\cite{mackenzie2017head}, with some programs using it to assess surgeons’ readiness for independent practice~\cite{miskovic2011development}. However, without structured guidance, video-based assessment alone is insufficient for effective skill acquisition. The use of artificial intelligence (AI) for automated skill assessment has shown great promise ~\cite{patel2009coming,kirubarajan2022artificial}. While AI-driven assessment offers better scalability, like video-based methods, it often lacks the actionable feedback necessary to support meaningful skill improvement. 
Most AI solutions function as black boxes, offering overall skill ratings without providing clear, interpretable guidance \cite{liu2021towards,ghasemloonia2017surgical,wang2018deep,lavanchy2021automation,funke2019video}. To fully harness AI’s potential in surgical education, it is crucial to develop systems that not only assess performance but also deliver structured, user-specific insights that facilitate skill acquisition. 

Explainable AI (XAI) can address this gap by generating interpretable feedback based on measurable skill proxies—such as motion efficiency, tool trajectory, and hand stability—allowing trainees to understand their performance in a clinically relevant way~\cite{bkheet2023using}. 
By breaking down complex assessments into intuitive, actionable insights, XAI has the potential to bridge the gap between automated evaluation and real-world skill improvement. Despite its potential, limited empirical studies exist regarding whether AI-generated feedback meaningfully enhances skill development~\cite{yilmaz2024real}. A key question remains: Can XAI effectively accelerate the acquisition of surgical skills while maintaining trainee engagement and educational value? 
Furthermore, the impact of AI-driven feedback on human factors, such as how trainees perceive, interpret, and integrate automated assessments, continues to be an open area of research.

In this study, we develop a simulation-based training framework that utilizes XAI to deliver user-specific, interpretable feedback for surgical skill acquisition. Through a prospective user study, we evaluate how XAI-driven feedback affects skill improvement, cognitive load, and trainees’ perceptions of AI-assisted learning. By comparing this method to traditional expert demonstrations, we aim to determine whether explainable feedback offers a meaningful advantage in skill acquisition and enhancing the training experience. By aligning AI-driven feedback with principles of deliberate practice, our approach represents a step toward integrating AI-driven coaching tools into surgical training at scale.

\section{Methodology}

\subsection{Study design}
We design a two-arm randomized study to evaluate the effectiveness of different feedback types--XAI-generated vs traditional video-based--in teaching suturing skills in a simulation environment.
XAI-generated intrinsic feedback was designed to provide user-specific guidance to improve their performance by comparing expert best practices and contrasting their execution with the maximally different primitive suturing actions.
We compare this approach to traditional video-based coaching, where trainees learn through self-guided observation of expert demonstrations. 
We hypothesize that trainees provided with XAI feedback will show greater improvements in the targeted skill concepts and enhanced perceptions of AI-assisted learning.

The experimental task emulated wound closure procedures with four interrupted instrument-tied sutures using a needle driver, surgical forceps, and suture scissors in incisions made on a skin suturing board~\cite{goldbraikh2022using}. Each suturing board was divided into quadrants. To standardize incision orientation and suture placement, participants always worked in the top-left quadrant and placed the first suture at the furthest end of the incision. Video recordings were captured using an Intel RealSense D435i RGBD sensor, though only RGB data was used for processing. We recruited 12 medical students with previous exposure to suturing and randomly assigned them to one of the two feedback interventions.

\begin{figure}[t]
    \centering
    \includegraphics[width=1.0\textwidth]{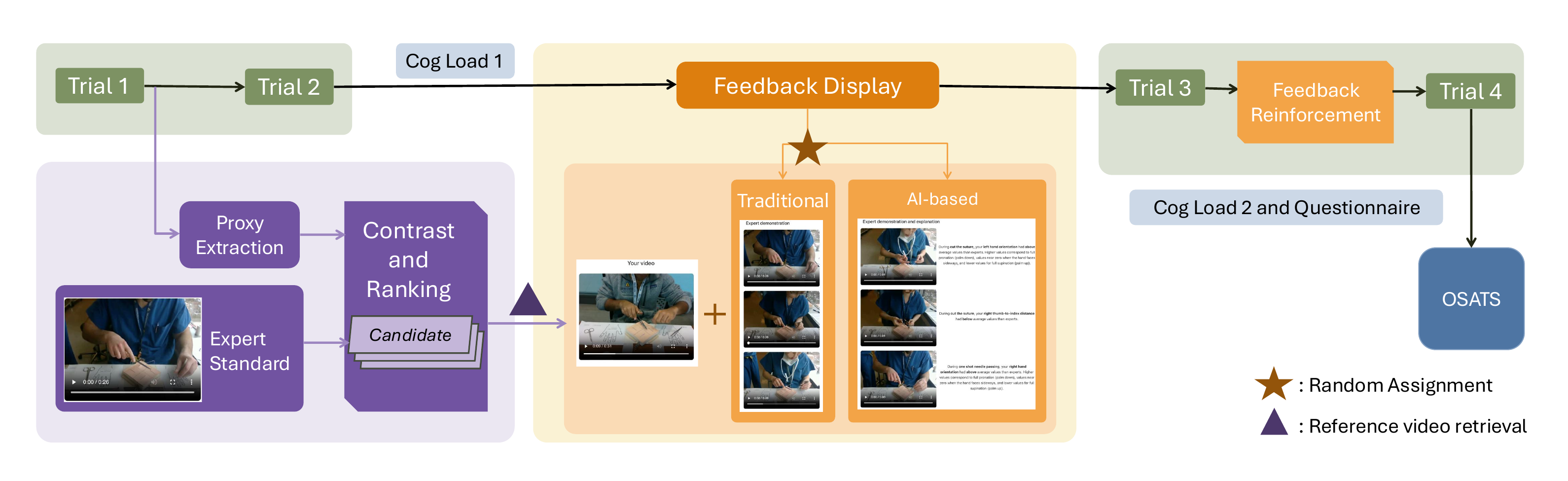} 
    \caption{Steps in the execution of the user study. The purple panel corresponds to the automated skill assessment and the orange one to the feedback interventions.}
    \label{fig:study_pipeline}
\end{figure}

In our study (Fig.~\ref{fig:study_pipeline}), participants completed four trials across two sessions, with a feedback intervention between sessions. The first session was followed by the cognitive load assessment and a break. We randomly assigned participants to a feedback type and gave a fixed time for feedback interpretation. 
In the second session, the same feedback was presented again before the last trial to reinforce its effect. This session concluded with another cognitive load assessment and feedback on perceptions. 
Last, participants viewed two of their trial videos (the second of each session) and rated their skills using a condensed global average rating scale (OSATS).    

\subsection{Feedback generation}

\noindent \textbf{Automated skill assessment:}
We employ automated skill assessment to analyze hand motion for fine-grained surgical performance analysis. Following the approach in~\cite{bkheet2023using}, we first perform gesture prediction based on the motion features extracted using object detection and 2D pose estimation. The YOLOX-S model detects hands and surgical tools. A pose estimation model predicts hand keypoints, which are then refined using post-processing techniques. These features serve as input for the MSTCN++ model, which classifies gestures into six categories: "No Gesture," "Needle Passing (G1)," "Pull The Suture (G2)," "Instrumental Tie (G3)," "Lay The Knot (G4)," and "Cut The Suture (G5)," based on the open surgery simulation dataset~\cite{goldbraikh2022using}. Here, we use the pre-trained models from~\cite{bkheet2023using}.
We then calculate the proxies for each detected gesture to quantify interpretable surgical action primitives. These proxies provide quantitative measures of hand pose characteristics relevant to skill assessment, using measurements previously identified as clinically relevant to analyze motion features~\cite{bkheet2023using}.

\begin{itemize}
    \item{\textbf{Hand Orientation (HO):}} This proxy measures how much the hand is turned. High values mean the palm is facing down (pronated), values near zero mean the hand is sideways, and low values mean the palm is facing up (supinated). Clinically, appropriate hand orientation is crucial for executing precise surgical movements. For instance, a slightly supinated hand can help achieve an optimal cutting angle when severing a suture~\cite{bkheet2023using}.
    \item{\textbf{Distance between Thumb and Index finger (DF):}} This proxy measures the space between the tips of the thumb and index finger and reflects different ways of holding tools and sutures. Surgeons typically secure the needle driver by resting it in the palm and grip sutures with their fingertips for precision, while novices rely on less refined grips that favor stability over accuracy~\cite{bkheet2023using}.
\end{itemize}

\noindent\textbf{Expert data collection:}
We collected expert data ($N=5$) under the same experimental setup to establish a benchmark for skill. Experts completed four trials, performing four sutures per trial. These samples were processed using the gesture prediction and proxy calculation pipeline described above.
To accurately calculate the proxies, we used the model-predicted hand poses and tools detection, and manually annotated surgical gestures. 
For each gesture, we computed the corresponding proxy values over time. Then, we average these values for each proxy-gesture pair $(c-g)$ with $c=\{\text{RHO, LHO, RDF, LDF}\}$ and ~$g=\{\text{G1, G2, G3, G4, G5}\}$, as $P = \frac{1}{n}\sum_{t=1}^n p_t $, where $p_t$ is the proxy value at time $t$. To establish a "standard" proxy-gesture pair representing typical expert performance, we combined values across experts over multiple trials:

\begin{equation}
    P_{ref} = \frac{1}{N\times T} \sum_{j=1}^{N} \sum_{i=1}^{T} P_{j,i}
\end{equation}

\noindent where, $N = 5$ experts, $T = 4$ trials, and $P_{j,i}$ is the average proxy value for expert $j$ at trial $i$. 

Using these expert-derived standard values, we selected video segments that best matched the standard proxy-gesture pairs.
We created two formats of the retrieved video to align with different feedback paradigms. For the traditional feedback condition, we provided a reference video showing a complete suture placement as a broad learning resource. For the XAI feedback, we curated video fragments, each focusing on a specific gesture related to the proxy measurement, serving as a more targeted exemplar of the proxy. 

\noindent \textbf{Feedback presentation through the user interface:}
We processed each participant's first trial ($P_{j,1}$) to provide real-time feedback intervention after the first session. To manage the information load during the intervention, we selected a subset of proxy-gesture combinations. For each participant, we first calculated the absolute difference between the average proxy value for each gesture and the corresponding expert standard, namely $P_{j,1} - P_{ref}$ for each $(c,g)$ pair, following the strategy to compare a new sample to the average of experts~\cite{bkheet2023using}. We then selected the top three differences with the largest deviations. We developed a custom user interface to present this feedback, including a self-video recording for comparison and recall, specifically the participant’s second suture from the first trial that was used to generate the feedback. For both interventions, we presented expert demonstrations for the top three proxy-gesture pairs, i.e., either a full suture demonstration or a video clip focusing on the specific gesture related to the proxy. The explainable feedback intervention included explicit guidance next to each expert video, following this template: "During [\emph{gesture}], your [\emph{hand}] [\emph{proxy name}] had [\emph{relative position}] average values than experts."

\subsection{Measures, data analysis and statistics}

We report proxy values for the proxy-gesture pairs (included in the feedback intervention) for the $2^{nd}$ and $4^{th}$ trials.
To accurately capture potential changes from the feedback intervention, we manually annotated the gestures for all sutures in these trials.
First, to quantify the performance gap with experts, we calculate the relative difference at the $i^{th}$ trial for the $j^{th}$ participant as: 
$S_{j,i} = |P_{j,i}-P_{ref}|/{P_{ref}}$.
We then measured improvement by calculating the difference between performance gaps in the second and fourth trials.
Second, we quantify the relative change of the proxy values before and after the intervention by calculating the difference in the proxy measurements for each participant. 
Cognitive load was assessed using the NASA TLX questionnaire (mental and physical demand, performance, effort, and frustration). Participants also rated their confidence in performing sutures on a ten-point scale and feedback usefulness and their understanding using a five-point Likert scale~\cite{alameddine2018video}. 
We displayed two video clips from each participant for a self-evaluation using a simplified OSATS.
For ordinal data from the survey responses, we used a non-parametric two-way Analysis of Variance (ANOVA) for session and feedback type comparisons, and independent t-tests (or a Mann–Whitney U test if assumptions are unmet) for continuous variables to be compared across feedback types.

\section{Results}

\textbf{a) Change in the relative gap between participants' proxy measurements and expert standards with different feedback:}

\begin{figure}[t]
    \centering
    \includegraphics[width=0.75\textwidth]{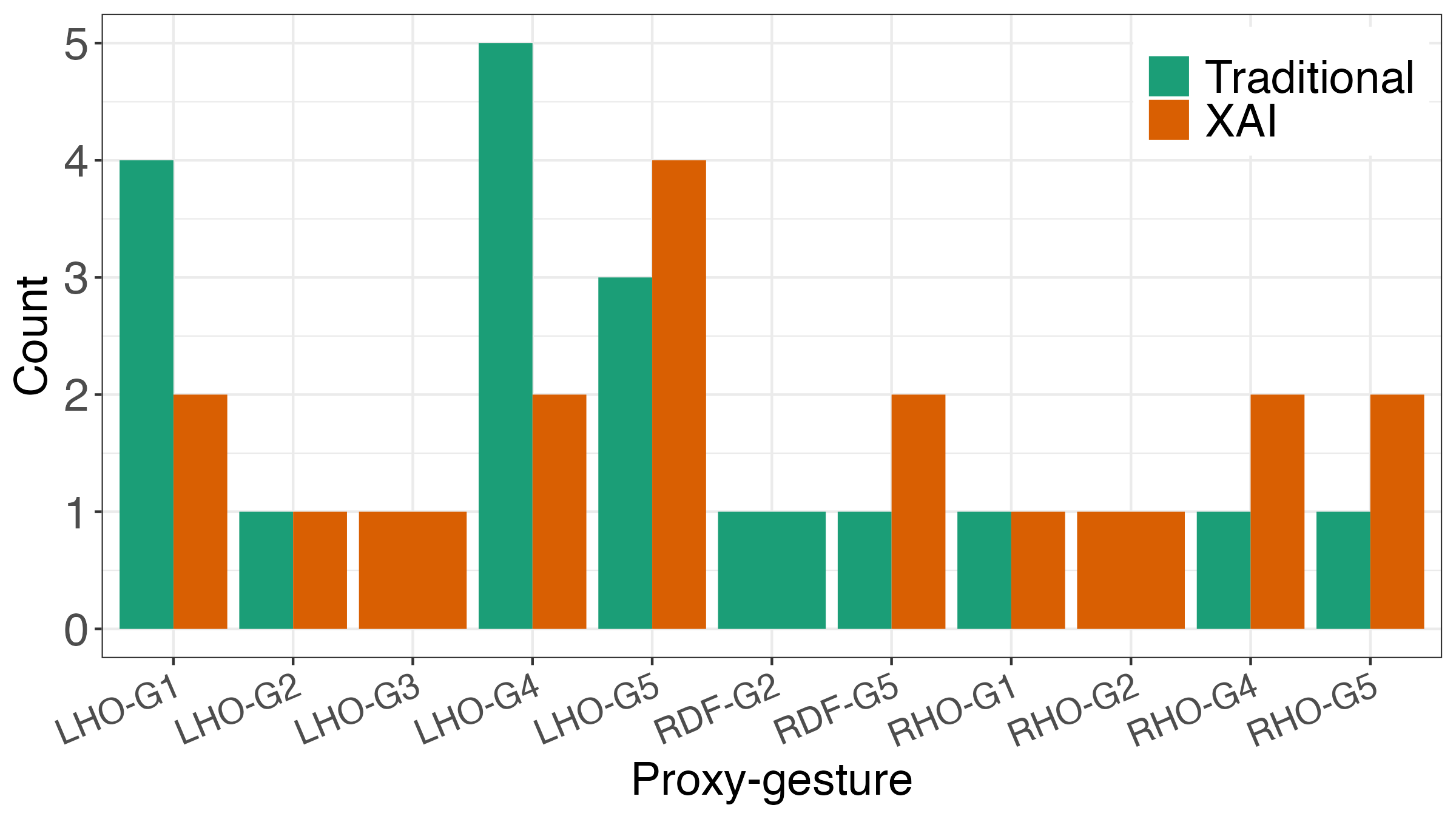} 
    \caption{Distribution of the top-three proxy-gesture pairs that had the largest deviations in the first trial and were presented to participants at each group.}
    \label{fig:hist_proxies}
\end{figure}

Fig.~\ref{fig:hist_proxies} shows the distribution of the top-three proxy-gesture pairs that had the largest deviations in the first trial and were used for the feedback intervention, representing the specific proxy measurements on which we quantified relative gaps.
We measured the difference in the relative gap for the top-three concepts before and after the feedback intervention, and averaged those differences into a single measurement per participant. The traditional feedback group showed an average change of 0.55 ($SD=0.94, CI: [-0.44, 1.54]$), while the XAI feedback group showed an average change of -0.46 ($SD=1.77, CI: [-2.32, 1.40]$). The positive change suggests that participants in the traditional group moved further from expert standards, while the negative value shows that XAI feedback helped participants move closer. However, the difference was not significant ($t(7.6)=1.23, p=0.254$).

\noindent \textbf{b) Change in proxy measurements before and after different feedback interventions:}

%
For each of the top-three ranked proxy-gesture pairs, we calculated the relative change before and after presenting the feedback and averaged these change values for each participant. On average, the relative change in proxy measurements was larger in the traditional group ($M=0.54, SD=0.60, CI: [-0.09, 1.17]$) than the XAI group ($M=0.37, SD=0.29, CI: [0.02, 0.73]$), but not significantly ($W=16, p=0.927$). 
Unlike the traditional group, the average relative change in the XAI group was significantly different from zero, indicating measurable proxy value changes after receiving explainable feedback.

\noindent \textbf{c) Impact of feedback interventions on subjective measurements:}
%
%
The feedback type did not significantly affect participants' cognitive load ($F(1,10)=0.40, p=0.540$), with an average of 4.07 ($SD=1.19$) and 4.50 ($SD=0.86$) in the traditional and XAI groups, respectively. 
However, session progression showed a significant effect ($F(1,10)=7.80, p=0.019$), with cognitive load decreasing from before ($M=4.63, SD=1.13$) to after the feedback intervention ($M=3.93, SD=0.84$), regardless of the information presented. No significant interaction effect between feedback type and session was observed ($F(1,10)=0.09, p=0.770$).
For perceived usefulness (Cronbach's $\alpha=0.92$), participants in the XAI group rated it slightly higher ($M=3.67, SD=0.98$) than those in the traditional group ($M=3.42, SD=0.67$). However, this difference was not significant ($W=15, p=0.677$). For perceived understanding (Cronbach's $\alpha=0.74$), both groups reported the same average rating ($M=3.58$), but variability was higher in the XAI group ($SD=0.86$) compared to the traditional group ($SD=0.38$). This comparison was also non-significant ($W=15, p=0.672$).

\begin{figure}[t]
    \centering
    \includegraphics[width=0.42\textwidth]{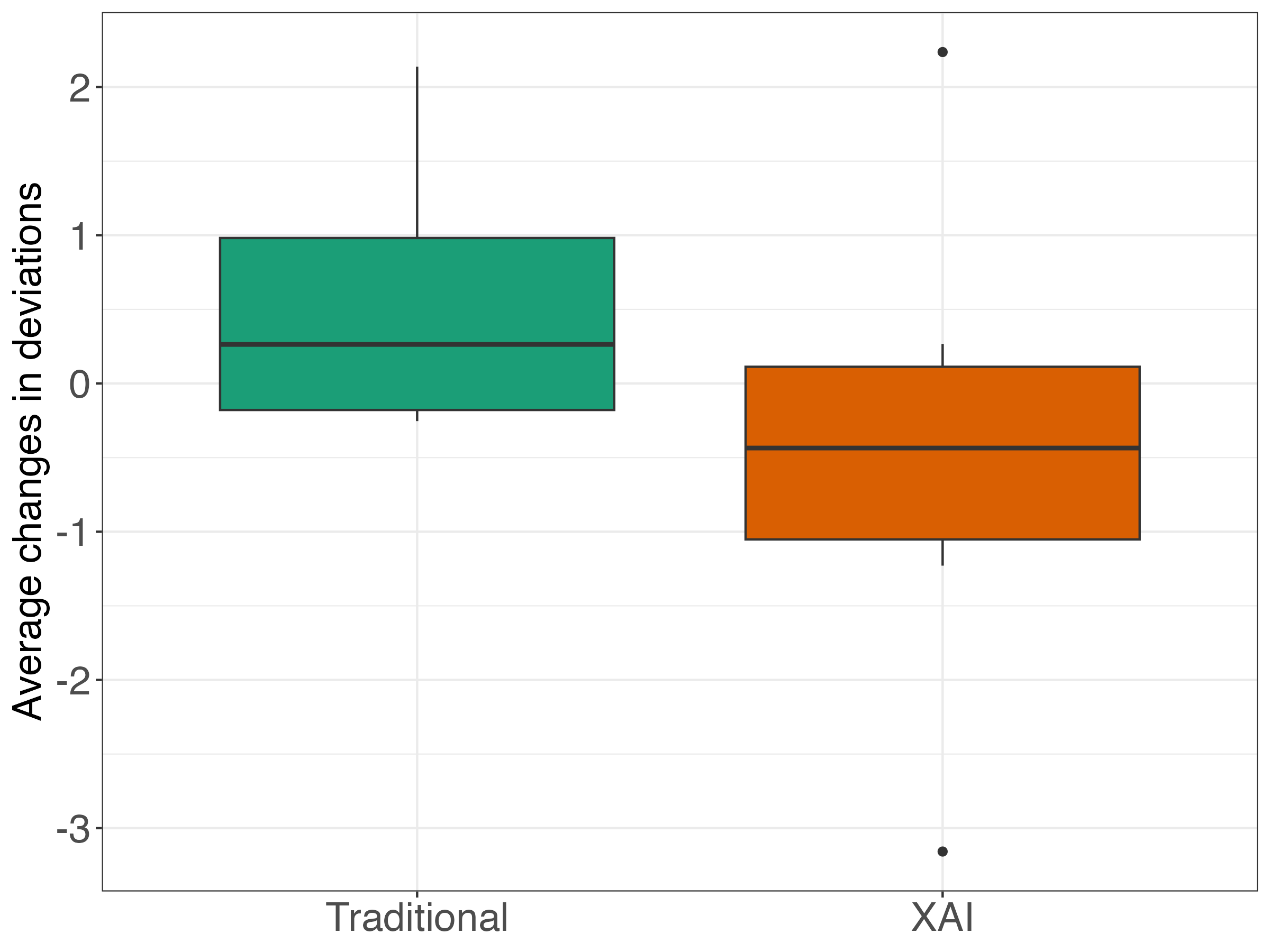} 
    \includegraphics[width=0.42\textwidth]{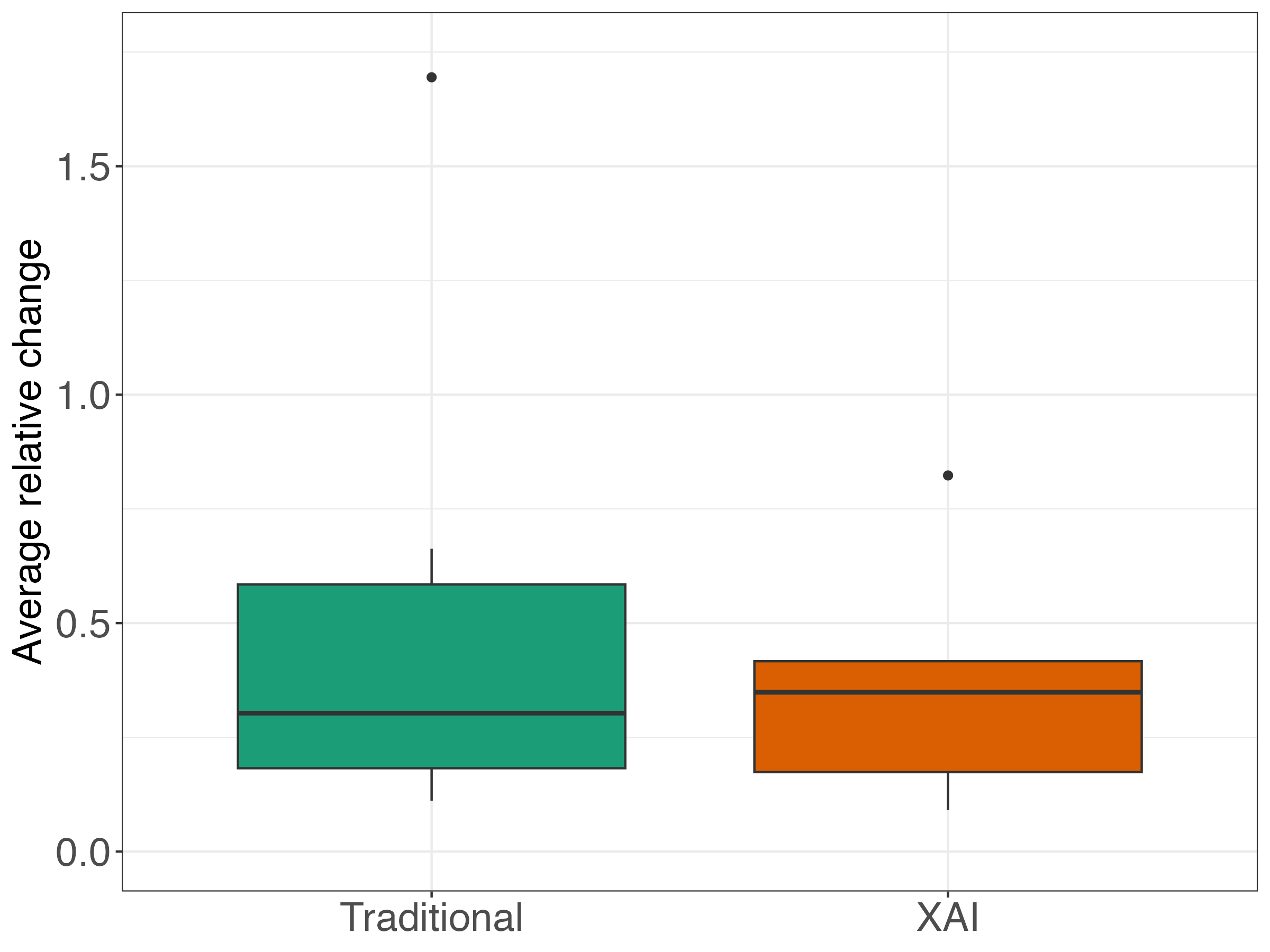} 
    \caption{Box and whisker plots for average change in deviations with respect to experts and average relative change in proxy values for the top three.}
    \label{fig:quant_results}
\end{figure}



Feedback type had no significant effect on participants' confidence ($F(1,10)=0.24, p=0.632$), with the traditional feedback group reporting slightly higher confidence ($M=4.33, SD=1.92$) than the XAI feedback group ($M=4.17, SD=1.27$). 
Regardless of feedback type, participants felt significantly more confident after the training intervention ($M=4.92, SD=1.44$) than before ($M=3.58, SD=1.50$), $F(1,10)=11.72, p=0.007$. The interaction effect was not significant ($F(1,10)=0.13, p=0.725$). 
Among the perceived competence constructs, only instrument handling showed a significant effect of session progression ($F(1,10)=5.79,p=0.037$), with ratings increasing from before ($M=2.50, SD=0.80$) to after ($M=3.00, SD=0.85$). 
There was no significant effect of feedback type on this competence ($F(1,10)=0.24, P=0.633$) with participants in the traditional feedback group rating on average 2.58 ($SD=1.00$) and in the XAI group reporting a slightly higher self-assessment ($M=2.92, SD=0.67$). The interaction effect was not significant ($F(1,10)=0.24, p=0.632$).

\begin{figure}[!t]
    \centering
    \includegraphics[width=0.75\textwidth]{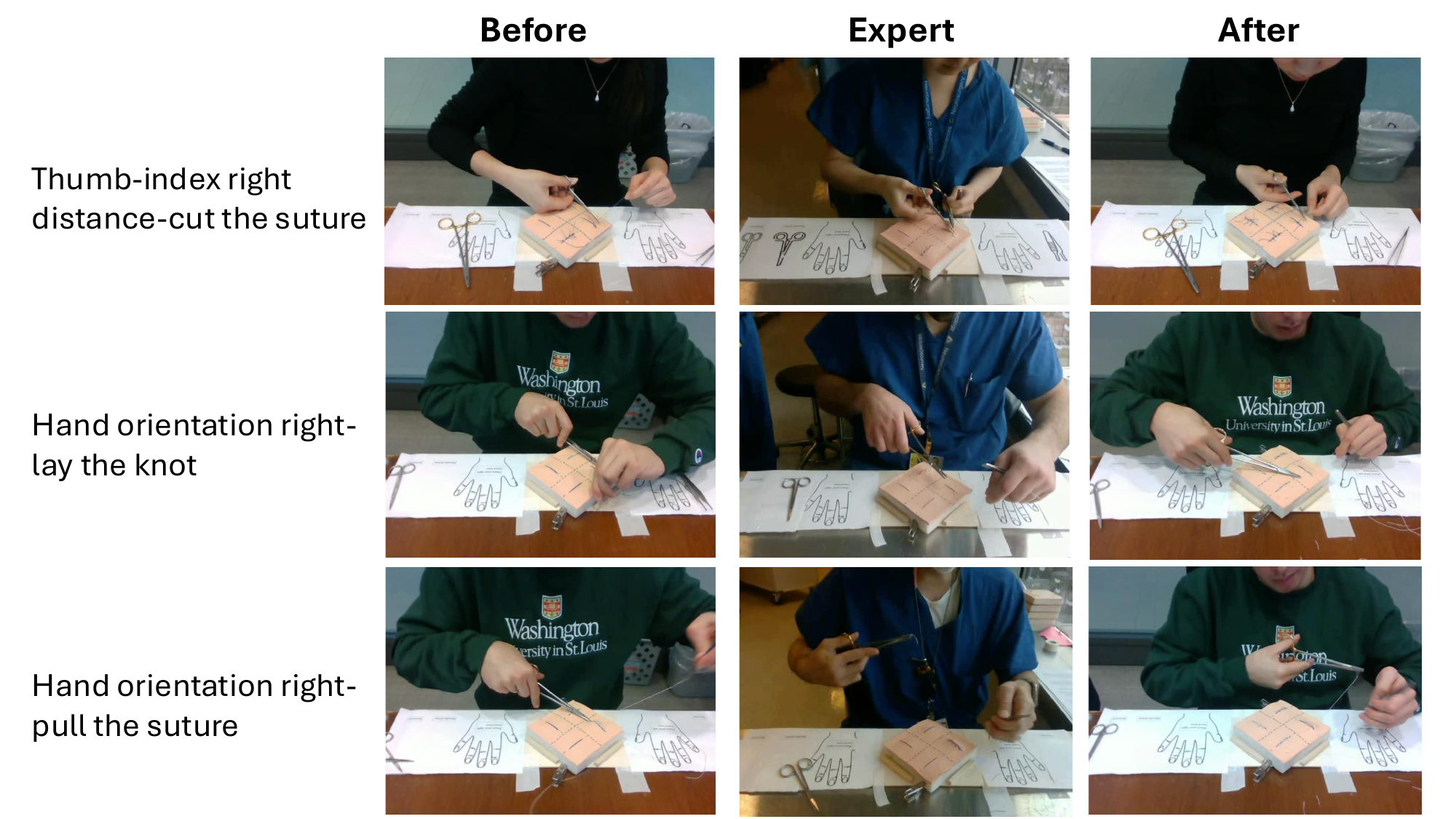} 
    \caption{Video clips from participants in the XAI group that showed less deviation from expert standards after receiving explained expert feedback. }
    \label{fig:qualitative_examples}
\end{figure}

\noindent \textbf{d) Qualitative results:}
Given the promising results of the XAI feedback intervention in moving participants closer to expert standards, we manually reviewed cases that showed notable improvements in reducing relative gaps among the top three concepts. Fig.~\ref{fig:qualitative_examples} shows samples from video clips of participants before and after feedback, alongside expert demonstrations for each proxy-gesture pair. 
In the first example, the average RDF was below the experts' average. After receiving XAI feedback, the participant adjusted closer to the experts' standards, resulting in a greater DF.
The second and third (rows) examples illustrate the RHO during different gestures: laying the knot and pulling the sutures. The expert clip shows that a less pronated hand position is needed, requiring a reduction in the participants' proxy measurement. Following the intervention, the participant successfully reduced their average proxy values and began using the needle driver to pull the suture, aligning more closely with the expert's technique.

\section{Discussion and Conclusion}
We present a novel implementation of AI-assisted methods for surgical skill training, with XAI providing near real-time personalized and interpretable feedback.  
Our initial findings did not demonstrate the superiority of XAI-generated feedback over traditional training methods in aligning performance with expert standards or shaping participant perceptions using statistical tests. We believe that this is for various reasons, that we will discuss separately below. 
Regardless, the large variations we observed in our measurements, compounded with the relatively small sample size, we believe that our results have to be interpreted with care.  Furthermore, while we used averaged proxy values for aggregating an individual's practice, the underlying concepts may exhibit temporal variation within each gesture, suggesting that temporal analyses could provide more meaningful comparisons between expert and novice performance.
Nevertheless, we did find positive trends both in quantitative and qualitative analyses, which we believe warrants further investigation of these XAI-based interventions in skill training.

The substantial deviations from expert standards indicate potential areas for improvement; however, the skillset of our participants with little surgical exposure may affect learning and their capacity to benefit from the intervention. The value of advanced feedback for transforming novices into experts appears limited when learners are still mastering fundamental skills and four trials may be insufficient to observe meaningful changes.
Notably, participants in both feedback groups reported similar increases in confidence over time and comparable feedback usefulness, suggesting that early-stage learners may focus more on gaining general task familiarity rather than responding differently to specific feedback formats. 
While participants may gain insights from feedback interventions, as shown in the qualitative examples with the desirable effect of XAI feedback, translating these insights into improved practice remains challenging for novices.

We identified methodological constraints in our intervention design that may have affected its effectiveness. The brief, single-trial feedback intervention was limited by real-time processing requirements and study duration. 
Moreover, our automated skill assessment uses existing out-of-the-box models~\cite{bkheet2023using}, pre-trained with its own experts and novice samples. Variations in surgical environments, lighting, or tool occlusions may affect detection, leading to gesture misclassification and unreliable proxies. 
The proxy calculations use pixel coordinates without depth information, potentially introducing errors in motion analysis, and the frontal view limits the capture of hand movements and tool interactions. 
Future implementations would benefit from adaptable models with improved robustness, depth estimation capabilities, and multi-view analysis to enhance spatial accuracy and provide more reliable feedback for effective skill assessment.

Incorporating understandable, task-specific feedback mechanisms with AI can automate the guidance of trainees toward expert-level performance. By identifying which aspects of AI-assisted learning support skill acquisition more effectively, we can iteratively improve these interventions to better align with the needs of trainees and develop human-centered AI for surgical education.

%
%
%
\bibliographystyle{splncs04}
\bibliography{references}

\subsubsection*{Acknowledgements}
This study was funded by the Johns Hopkins DELTA Grant IO 80061108 and the Link Foundation Fellowship in Modeling, Simulation, and Training. 

\end{document}